\documentstyle[12pt,epsfig]{article}
\def\lsim{\mathrel{\raise3pt\hbox to 8pt{\raise -6pt\hbox{$\sim$}\hss{$<$}}}}

\def\haf{\textstyle{1\over2}}
\def\thaf{\textstyle{3\over2}}

\def\ttrd{\textstyle{2\over3}}
\def\ftrd{\textstyle{4\over3}}
\def\minus{\mbox{$-$}}

\newcommand{\vx}{\vec{x}}

\newcommand{\vy}{\vec{y}}
\newcommand{\vz}{\vec{z}}

\newcommand{\vq}{\vec{\, q}}

\newcommand{\vJ}{\vec{J}}
\newcommand{\vL}{\vec{L}}
\newcommand{\vS}{\vec{S}}
\newcommand{\vsig}{\vec{\sigma}}
\newcommand{\vtau}{\vec{\tau}}

\newcommand{\vmu}{\vec{\mu}}
\newcommand{\vp}{\vec{p}}
\newcommand{\vnabla}{\vec{\nabla}}

\newcommand{\hxij}{\hat{x}_{i j}}

\newcommand{\bbC}{\bar{\! \bar{C}}}
\newskip\humongous \humongous=0pt plus 1000pt minus 1000pt
\def\caja{\mathsurround=0pt}

\newif\ifdtup
\def\panorama{\global\dtuptrue \openup1\jot \caja
        \everycr{\noalign{\ifdtup \global\dtupfalse
        \vskip-\lineskiplimit \vskip\normallineskiplimit
        \else \penalty\interdisplaylinepenalty \fi}}}
\def\eqalignno#1{\panorama \tabskip=\humongous
        \halign to\displaywidth{\hfil$\displaystyle{##}$
        \tabskip=0pt&$\displaystyle{{}##}$\hfil
        \tabskip=\humongous&\llap{$##$}\tabskip=0pt
        \crcr#1\crcr}}
\begin{document}
\vspace*{-0.6in}
\hfill \fbox{\parbox[t]{1.12in}{LA-UR-04-6214}}\hspace*{0.35in}
\vspace*{0.0in}

\begin{center}

{\Large {\bf The Nuclear Physics of Hyperfine Structure in Hydrogenic Atoms}}\\
\vspace*{0.4in}
{\bf J.\ L.\ Friar} \\
{\it Theoretical Division,
Los Alamos National Laboratory \\
Los Alamos, NM  87545} \\
\vspace*{0.20in}
and \\
\vspace*{0.10in}
{\bf G.\ L.\ Payne}\\
{\it Dept. of Physics and Astronomy\\
Univ. of Iowa\\
Iowa City, IA 52242}
\end{center}

\begin{abstract}
The theory of QED corrections to hyperfine structure in light hydrogenic atoms
and ions has recently advanced to the point that the uncertainty of these
corrections is much smaller than 1 part per million (ppm), while the experiments
are even more accurate. The difference of the experimental results and the
corresponding QED theory is due to nuclear effects, which are primarily the
result of the finite nuclear charge and magnetization distributions. This
difference varies from tens to hundreds of ppm. We have calculated the
dominant nuclear component of the 1s hyperfine interval for deuterium, tritium
and singly ionized helium, using a unified approach with modern
second-generation potentials. The calculated nuclear corrections are within 3\%
of the experimental values for deuterium and tritium, but are roughly 20\%
discrepant for helium. The nuclear corrections for the trinucleon systems can be
qualitatively understood by invoking SU(4) symmetry.
\end{abstract}
\pagebreak
\section{Introduction}

Until very recently hyperfine splittings in light hydrogenic atoms were the most
precisely measured atomic transitions.  Theoretical predictions based on QED are
less accurate, but have improved considerably in recent years.  Non-recoil and
non-nuclear contributions\cite{eides01,savely02} are known through order
$\alpha^3 E_{\rm F}$, where $E_{\rm F}$ is the Fermi hyperfine energy (viz., the
leading-order contribution) and $\alpha$ is the fine-structure constant. 
Because the hadronic scales for recoil and certain types of nuclear corrections
are the same, recoil corrections are treated on the same footing as nuclear
corrections\cite{eides01}, and we will call both types ``nuclear corrections.''
Uncalculated QED terms of order $\alpha^4 E_{\rm F}$ in light atoms are almost
certainly smaller than .1 ppm., while the experimental errors are smaller still.
This provides us with the unprecedented opportunity to study nuclear effects in
the hyperfine structure (hfs) of light hydrogenic atoms, which range in size 
from tens to hundreds of ppm. We will restrict ourselves to hydrogenic
s-states, because these states maximize nuclear effects.

\begin{table}[htb]
\centering

\caption{Difference between hyperfine experiments and QED hyperfine calculations
for the $n\underline{\rm th}$ s-state of light hydrogenic atoms times $n^3$,
expressed as parts per million of the Fermi energy. This difference is
interpreted as nuclear contributions to the hyperfine
splitting\protect\cite{savely02}. A negative entry indicates that the
theoretical prediction without nuclear corrections is too large.}
\vspace*{0.1in}
\begin{tabular}{|l || c c c c|}
\multicolumn{5}{c}{$n^3 (E_{\rm hfs}^{\rm exp} - 
E_{\rm hfs}^{\rm QED})/E_{\rm F}\, {\rm (ppm)}$} \\ \noalign{\smallskip} \hline
 State & H & $^2$H & $^3$H & $^3$He$^+$ \rule{0in}{2.5ex}\\ \hline \hline
 1s    & $-$33   & 138    & $-$38       & $-$212        \\ \hline
 2s    & $-$33   & 137    &\mbox{$--$}   & $-$211        \\ \hline 
\end{tabular}
\end{table}

Table 1 is an updated version of the corresponding table in Ref.\cite{savely02}.
Because nuclear effects have a much shorter range than atomic scales, we expect
the splittings in the $n\underline{\rm th}$ s-state to be proportional to
$|\phi_n(0)|^2 \sim 1/n^3$, where $\phi_n (r)$ is the non-relativistic wave
function of the electron. Forming the fractional differences (in parts per
million) between $E^{\rm exp}_{\rm hfs}$ and $E^{\rm QED}_{\rm hfs}$ (times
$n^3$) leads to the tabulated results. As we stated above, these large
differences reflect neither experimental errors nor uncertainties in the QED
calculations, but rather the large nuclear contributions to hfs.

In order to perform a tractable calculation it is necessary to restrict the
scope, while at the same time incorporating the dominant physics. To accomplish
this we borrow a technique from chiral perturbation theory ($\chi$PT, the
effective field theory for nuclei based on QCD). This technique, called power
counting, is the organizing principle of $\chi$PT and allows one to perform
systematic expansions\cite{weinberg} in powers of a small parameter,
$(Q/\Lambda)$, where $Q$ is a typical nuclear momentum scale that can be taken
to be roughly the pion mass ($m_\pi \sim$ 140 MeV), and $\Lambda$ is the
large-mass QCD scale ($\sim$ 1 GeV) typical of QCD bound states such as the
nucleon, heavy mesons, nucleon resonances, etc. We also note that $1/Q$
specifies a typical correlation length (and a reasonable nearest-neighbor
distance) in light nuclei ($\sim$ 1.4 fm)\cite{primer}. This expansion in 
powers of $(Q/\Lambda \sim 0.1-0.15)$ should converge moderately well. In this
work we restrict ourselves to leading order in this expansion, and this
restriction eliminates nuclear corrections of relativistic order, which are
subleading and exceptionally complicated because of the complexity of the
nuclear force\cite{dhg,yulik}.

In processes that involve virtual excitation of intermediate nuclear states
(each with its own energy, $E_N$, relative to the ground-state energy, $E_0$)
the excitation energy $(\omega_N = E_N - E_0)$ is of order $Q^2/\Lambda$ and
typically is a correction to the leading order\cite{primer}. Consistency 
therefore demands that we drop such terms.  The nuclear recoil energy scales
like $Q^2/M$, where $M$ is the nucleon mass, and can also be dropped. These are
very considerable simplifications in constructing the nuclear contribution to
hfs, which we present in the next section.

\section{Nuclear Contributions to Hyperfine Structure}

The hyperfine interactions that interest us are simple (effective) couplings of
the electron spin to the nuclear (ground-state) spin: $\vsig \cdot \vS$, where
$\vsig$ is the electron (Pauli) spin operator and $\vS$ is the nuclear spin
(total angular momentum) operator. Other couplings of the electron spin are
possible and either generate no hyperfine splitting, none in s-states, or
higher-order (in $\alpha$) contributions. 

The lowest-order (Fermi) hyperfine interaction is generated by the interaction
of the electron current with the magnetic dipole part (determined by $\vmu$) of
the nuclear current. A simple calculation gives the well-known
result\cite{eides01}
$$
E_{\rm F} = \frac{4 \pi \alpha \mu_N |\phi_n (0)|^2} {3 m_e}
\frac{\vsig \cdot \vS}{S} \, , \eqno(1)
$$
where $\langle S S | \mu_{\rm z} |S S \rangle \equiv \mu_N$ defines the nuclear
magnetic moment and $m_e$ is the electron mass. The factor of $(\vsig \cdot 
\vS)/S$ leads to a hyperfine splitting proportional to $(2\, S + 1)/S$.
All additional contributions will be measured as a fraction of this energy.

\begin{figure}[htb]
\epsfig{file=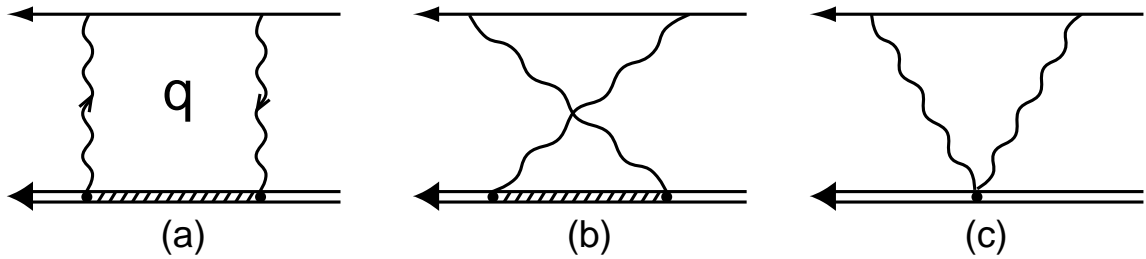,height=1.32in}
\caption{Nuclear Compton amplitude with direct (a), crossed (b), and seagull (c)
contributions illustrated. Single lines represent an electron, double lines a
nucleus, and shaded double lines depict a nuclear Green's function containing a
sum over nuclear states. The seagull vertex maintains gauge invariance. The 
four-momentum running through the loop is $q^{\mu}$.}
\end{figure}

Naively calculating the higher-order (in $\alpha$) corrections using only the 
nuclear magnetization distribution will fail. The atomic wave function 
is modified by the nuclear charge distribution in precisely the same region 
that the magnetization distribution is nonvanishing\cite{BW,zemach56}. It is 
therefore necessary to incorporate the complete set of second-order (in
perturbation theory in $\alpha$) processes shown in Fig~(1), which comprise the
nuclear Compton amplitude coupled to the electron Compton amplitude. Only the
forward-scattering part of this amplitude is required for the ${\cal O}
(\alpha)$ corrections to $E_{\rm F}$.  The resulting energy shift is then given
by
$$
\Delta E = i (4 \pi \alpha)^2 |\phi_n (0)|^2 \int \frac{d^4 q}{(2 \pi)^4}
\frac{t_{\mu \nu} (q)\, T^{\mu \nu} (q,-q)}
{(q^2 + i \epsilon)^2 (q^2 - 2 m_e q_0 + i \epsilon)}\, , \eqno(2)
$$
where $t^{\mu \nu}$ is the electron Compton amplitude and $T^{\mu \nu}$ is the
nuclear Compton amplitude, which is required to be gauge invariant.  The lepton
tensor can be decomposed into an irreducible spinor basis, and we can ignore odd
matrices and spin-independent components.

We also ignore (for now) terms that couple two currents together. It is easy to
show that since the nuclear current scales as $1/\Lambda$ (the conventional
components of the current have explicit factors of $1/M$), two of them should
scale as $1/\Lambda^2$ and generate higher-order (in $1/\Lambda$) terms.  This
leaves a single dominant term representing a charge-current correlation
$$
\Delta E = (4 \pi \alpha)^2 |\phi_n (0)|^2 \int \frac{d^4 q}{(2 \pi)^4}
\frac{(\vsig \times \vq)^m [T^{m 0} (q,-q) - T^{0 m} (q,-q)]}
{(q^2 + i \epsilon)^2 (q^2 - 2 m_e q_0 + i \epsilon)}\, . \eqno(3)
$$
The nuclear seagull terms $B^{0 m}(q,-q)$ and $B^{m 0}(q,-q)$ that are contained
as part of $T^{\mu \nu}$ in Fig.~(1c) are of relativistic order\cite{dhg} $(\sim
1/\Lambda^2)$ and can be dropped. Although the seagull terms $B^{m n}
(\vq,-\vq)$ are of non-relativistic order, they do not contribute to hfs because
of crossing symmetry. The explicit form for the remaining term in $T^{m 0}$
(suppressing the nuclear ground-state expectation value), which involves a
complete sum over intermediate states, $N$, is given by
$$
T^{m 0} (q,-q) = \sum_{N} \frac{J^m (-\vq)\, |N \rangle \langle N | \,
 \rho (\vq)}{q_0 - \omega_N + i\epsilon} + {\rm crossed\; term}
\, , \eqno(4)
$$
where $\vJ (\vq)$ and $\rho (\vq )$ are the Fourier transforms of the nuclear 
current and charge operators. The crossed term has the operator order reversed
and $q_0 \rightarrow - q_0$.

The limits $\omega_N \rightarrow 0$ and $m_e \rightarrow 0$ (both scales are 
small compared to the nuclear-size scale, $1/Q$) greatly simplify the
calculation, and this leads to
$$
\Delta E = i \,(4 \pi \alpha)^2 |\phi_n (0)|^2 \int \frac{d^3 q}{(2 \pi)^3}
\frac{(\vsig \times \vq)^m \{J^m (-\vq) , \rho (\vq)\}}
{\vq^6}\, , \eqno(5)
$$
which is infrared divergent.  Using
$$
J^m (-\vq) = \int d^3 y\, J^m (\vy) \exp{(-i \vq \cdot \vy)}\, , \eqno(5)
$$
and
$$
\rho (\vq) = \int d^3 x \, \rho (\vx) \exp{(i \vq \cdot \vx)}\, , \eqno(6)
$$
together with $\vz \equiv \vx - \vy$, and a lower-limit (infrared) $q-$cutoff, 
$\epsilon$, we find
$$
\Delta E = -8 \alpha^2 |\phi_n (0)|^2 \int d^3 x \int d^3 y \, \{ \rho (\vx), 
\vsig \cdot \vJ (\vy) \} \times \vnabla_z (\frac{1}{3 \epsilon^3} - 
\frac{z^2}{6 \epsilon} +\frac{\pi z^3}{48})\, , \eqno(7)
$$
where there is an implied (nuclear and atomic) expectation value. The constant
term does not contribute because of the derivative. The second term is only
logarithmically divergent when Siegert's theorem is applied\cite{MEC}, can be
shown to vanish in several limits (including the zero-range limit), and is 
consequently negligibly small\cite{big}. The last term is the one we are
seeking and was originally developed by Low\cite{d-th-1} in a limiting case for
the deuteron, after the basic concept was sketched by Bohr\cite{d-th-0}:
$$
\Delta E^{\rm Low}_{\rm hfs} = \frac{\pi \alpha^2 |\phi_n (0)|^2}{2} 
\int d^3 x \int d^3 y \, |\vx - \vy|\, \{ \rho (\vx), \vsig \cdot (\vx - \vy)
\times \vJ (\vy) \} \, + \cdots \, . \eqno(8)
$$
A more convenient representation can be
obtained by dividing both sides of this equation by the expression for the Fermi
hyperfine energy given by Eqn.~(1).  Since the Wigner-Eckart theorem guarantees
that Eqn.~(8) must be proportional to $\vsig\cdot\vS/S$ (which cancels in the
ratio), we arrive at a simple expression for the leading-order nuclear 
contribution to the hfs, which is one of our primary results.
$$
\Delta E^{\rm Low}_{\rm hfs}/E_{\rm F} = -2\, m_e \, \alpha \, \delta_{\rm Low}
\, , \eqno(9)
$$
where
$$
\delta_{\rm Low} = -\frac{3}{16 \mu_N} \int d^3 x \int d^3 y \, |\vx - \vy| \,
\{ \rho (\vx), ((\vx-\vy) \times \vJ (\vy))_{\rm z} \}  \, , \eqno(10)
$$
and a nuclear expectation value is required of the z (or ``3'') component of
this vector in the nuclear state with maximum azimuthal spin (i.e. $S_{\rm z} =
S$). The intrinsic size of the nuclear corrections is given by ($-2\, m_e\, 
\alpha R$) =  $-$38 ppm $[R/$fm], where $[R/$fm] is the value of the Low moment 
in Eqn.~(10) in units of fm.

\section{Nuclear Matrix Elements}

In order to evaluate Eqn.~(10) it is necessary to assume a form for the nuclear
charge and current operators. We have agreed to ignore terms of relativistic
order, and this eliminates all but the usual impulse-approximation (i.e.,
one-body) terms for the charge operator, which contains finite-size
distributions for the protons and neutrons. Although the latter contributions
are small, they have never been included in previous calculations and we will
gauge their importance by including them in our calculation. The nuclear current
operator can be written in terms of the dominant spin-magnetization current, the
convection current (motional current of charged particles) and meson-exchange
currents (MEC). Isoscalar MEC are of relativistic order\cite{dhg} and we ignore
them. Isovector MEC are larger, but don't contribute to the deuteron because it
is an isoscalar nucleus. Isovector MEC will contribute to the trinucleon 
systems, where the effect of this current on the isovector magnetic moment is
about the same size (15\%) as our expansion parameter\cite{mag-mom}. Because
parts of these currents (in particular the isobar part) are difficult to treat
quantitatively and because of their relative smallness, we will ignore MEC in
calculating the Low moments in this initial effort to understand hfs in the
trinucleon systems.

Each of the charge and current operators that we use are therefore given by sums
of one-body operators, and their resulting product in the Low-moment expression
in Eqn.~(10) can be written as a sum of one-body terms plus a sum of two-body
terms. Using a transparent notation for this decomposition ($\delta_{\rm Low} =
\delta^{(1)}_{\rm spin} + \delta^{(2)}_{\rm Low}$ and $\delta_{\rm Low}^{(2)} =
\delta_{\rm Low}^{\rm mag} + \delta_{\rm Low}^{\rm conv}$) we find that the
one-body term is given for all nuclei by the spin-magnetization current in the
form
$$
\delta_{\rm spin}^{(1)} = \langle r \rangle^{pp}_{(2)} \frac{\mu_p}{\mu_N}
\sum_{i=1}^A \left(\frac{1 + \tau^3_i}{2}\right) \sigma_i^{\rm z} +
\langle r \rangle^{nn}_{(2)} \frac{\mu_n}{\mu_N}
\sum_{i=1}^A \left(\frac{1 - \tau^3_i}{2}\right) \sigma_i^{\rm z}
\, , \eqno(11)
$$
where
$$
\langle r \rangle^{pp}_{(2)} = \int d^3 x \int d^3 y \, \rho^p_{ch} (x) \,
\rho^p_M (y) |\vx - \vy| = 1.086(12) {\, \rm fm} \,  \eqno(12a)
$$
and
$$
\langle r \rangle^{nn}_{(2)} = \int d^3 x \int d^3 y \, \rho^n_{ch} (x) \,
\rho^n_M (y) |\vx - \vy| \,  \eqno(12b)
$$
determine the proton and neutron parts, respectively, of the one-body current.
The quantities $\vtau_i$ and $\vsig_i$ are the (Pauli) isospin and spin
operators for the i\underline{th} nucleon, $\rho^p_{ch}$, $\rho^p_M$, and
$\rho^n_M$ are the proton charge and magnetic densities and the neutron magnetic
density (normalized to 1), while $\rho^n_{ch}$ is the neutron charge density
(normalized to 0). Note that the quantities $\langle r \rangle^{pp}_{(2)}$ and
$\langle r \rangle^{nn}_{(2)}$ are the proton and neutron Zemach
moments\cite{zemach56, pert}, and we have listed in Eqn.~(12a) the value of the
proton Zemach moment recently determined directly from the electron-scattering
data for the proton\cite{ingo04} (the neutron has not yet been evaluated). In
numerical work described below we will use simple forms for the neutron and
proton form factors:  a dipole form for the proton charge and magnetic form
factors and the neutron magnetic form factor $(F_D (q^2) = \frac{1}{(1 +
q^2/\beta^2)^2})$ and a modified Galster\cite{galster} form for the neutron
charge form factor $(F_G = \frac{\lambda q^2}{(1 + q^2/\beta^2)^3}$, where
$\lambda = 0.0190$ fm$^2$).  To incorporate into our calculations the numerical
value given by Eqn.~(12a) we use $\beta = 4.029$ fm$^{-1}$, which reproduces
that value for the dipole case. With this $\beta$ the neutron Zemach moment has
the value $-0.042$ fm, which is a very small correction to the proton. In Low's
original work the nucleon Zemach moments were ignored (at that time no
information existed that they were significant).

The spin-isospin operators in Eqn.~(11) are generators of SU(4) symmetry, and it
is conventional\cite{su4} to decompose the wave functions of light nuclei
with respect to that symmetry, which plays a significant role in understanding
those systems. The dominant wave function component for the trinucleon systems
(the S-state $\sim 90\%$) is the product of a completely symmetric space wave
function with a completely antisymmetric spin-isospin wave function. The next
most important component is the D-state ($\sim 9\%$), and  we will ignore the
small remaining components for simplicity in the following discussion\cite{big}.
Treating only the proton term for the moment, we find the expectation value of
$\sum_{i=1}^A \left(\frac{1 + \tau^3_i}{2}\right) \sigma_i^{\rm z}$ to be
$S_{\bf z} (1-\thaf P_D)$ for the deuteron, $2 S_{\bf z} (1- \ftrd P_D)$ for the
triton, and $2 S_{\bf z} (-\ttrd P_D)$ for $^3$He. The D-wave components tend to
have the spin and orbital components anti-aligned, and this accounts for the
sign of the $P_D$ term. In $^3$He the two protons tend to have their spins
oppositely aligned, which accounts for the small $^3$He Low moment. In the limit
of exact SU(4) symmetry only the S-state contributes and the $^3$He one-body
term vanishes.

There are three types of two-body Low moments: S-wave spin-magnetization terms,
D-wave spin-magnetization terms, and (largely) D-wave convection current terms.
For each of these types there are contributions from two protons, or from one
proton and one neutron, or from two neutrons. We keep all terms but the
convection current contribution from two neutrons. The resulting nuclear
operators are
$$\eqalignno{
\delta_{\rm Low}^{\rm mag} =& \frac{1}{\mu_N}      \sum_{i \neq j}^A
\left( \vsig_j C_{i j} (x_{i j}) - \frac{1}{8} \bbC_{i j} (x_{i j})
(3 \vsig_j \cdot \hxij \hxij - \vsig_j)\right)_{\rm z} \, , &(13a) \cr
\delta_{\rm Low}^{\rm conv} =& \frac{3}{16 M \mu_N} \sum_{i \neq j}^A
\bar{C}_{i j} (x_{i j}) \vL_{i j}^{\rm z} \, , &(13b) \cr}
$$
where $\vL_{ij} = \vx_{ij} \times (\vp_i - \vp_j)$, $\vx_{ij} = \vx_i - \vx_j$,
$\vx_i$ is the coordinate of nucleon $i$, and $\vp_i$ is the momentum of nucleon
$i$. For simplicity we have not decomposed the radial (and isospin-dependent)
functions $C_{i j}$, $\bar{C}_{i j}$, and $\bbC_{i j}$ according to the types of
nucleon that contribute. Explicit forms for these functions are given in
Ref.~\cite{big}. In the limit of pointlike nucleons the radial part of each 
function becomes simply $x_{i j}$.

\section{Results and Discussion}

The proton hfs has been discussed in detail recently\cite{savely02,ingo04} and
we have nothing additional to add. The recently evaluated proton Zemach moment
was discussed in the text, and it leads to a $-$58.2(6) kHz contribution to the
hydrogen hfs, which equals $-$41.0(5) ppm. When added to the usual QED and 
recoil corrections\cite{eides01,savely02,ingo04} there is a 3.2(5) ppm
discrepancy with experiment, which can be attributed to hadronic 
polarization\cite{hughes83,faustov02} and (possibly) uncalculated recoil 
corrections.

The deuterium, tritium, and $^3$He$^+$ results were calculated using the
(second-generation) AV18 potential\cite{av18}, together with (for $^3$H and
$^3$He) an additional TM$^{\prime}$ three-nucleon force\cite{TM} whose
short-range cutoff parameter had been adjusted to provide the correct binding
energies. Individual one-body (labelled ``nucleon'') and two-body (labelled
``Low'') terms are tabulated together with their total in Table 2. The
(approximate) SU(4) symmetry that dominates light nuclei\cite{big,su4} provides
an explanation for the relative sizes of the entries in this table, which we 
discuss next. Note that $^3$He (which has proton number $Z=2$) is uniformly
enhanced by a factor of $Z^3 = 8$ contained in $|\phi_n(0)|^2$ in Eqn.~(8).

\begin{table}[htb]
\centering

\caption{Nuclear corrections to 1s hyperfine structure in light hydrogenic
atoms. ``Nucleon'' refers to the one-body part, and ``Low'' refers to the
correlation or two-body term, and ``Total'' refers to their sum. All entries are
given in kHz.}
\vspace*{0.1in}
\begin{tabular}{|c||ccc||ccc||ccc|}
\hline
\multicolumn{1}{|c||}{H}& \multicolumn{3}{|c||}{$^2$H}
& \multicolumn{3}{|c||}{$^3$H}&\multicolumn{3}{|c|}{$^3$He$^+$}\\ \hline
Nucleon&Nucleon&Low&Total&Nucleon&Low&Total&Nucleon&Low&Total\\ \hline
$-$58.2(6)&$-$41.1&87.3&46.2&$-$50.6&$-$9.6&$-$60.1&13.9&1428&1442\\
\hline
\end{tabular}
\end{table}

We expect (and verify) that interactions driven by the charge of the neutron
will be significantly suppressed because the neutron is overall neutral. The
neutron Zemach moment, for example, is $\sim$ -- 4\% of that of the proton, and
this greatly suppresses the neutron's contribution to the one-body term. The two
protons in $^3$He have their spins anti-aligned in the SU(4) limit, and this
cancellation leads to a small net result for the one-body part. The protons in H
and $^3$H make comparable one-body contributions, since the proton in $^3$H
carries the entire spin in the SU(4) limit.

The two-body terms that couple the neutron charge and the proton magnetic moment
are suppressed for the reason discussed above. In addition the two-body
convection current has no contribution from the dominant S-state and is
therefore negligible for hydrogen isotopes, where one (charged) nucleon must be
a neutron. Only for the two protons in $^3$He is this interaction
non-negligible, but still small. In $^3$H the neutron spins are anti-aligned in
the SU(4) limit, which suppresses the normally dominant proton charge -- neutron
magnetic moment contribution to about 20\% of the one-body part. In $^3$He those
terms add for the two protons, leading to an even larger result. Thus the
qualitative features of the results in Table~2 can be understood in terms of
(approximate) SU(4) symmetry. We quantify these qualitative observations in the
following paragraph.

The neutron's contribution to the nucleon one-body term is very small, except 
for $^3$He where it is $\sim -40$\% of the very suppressed proton contribution.
In deuterium the largest correction to the dominant proton charge -- neutron
magnetic moment scalar interaction (the first of the two terms in Eqn.~(13a))
is the corresponding tensor interaction (the second term in Eqn.~(13a)) and
amounts to only 4\%. In tritium the proton charge -- neutron magnetic moment 
scalar interaction is suppressed by SU(4) symmetry, which enhances the
relative contribution of the corresponding tensor term ($\sim$ 30\%) and of
the scalar neutron charge -- proton magnetic moment term ($\sim$ 20\%). The
convection current contribution to both of these hydrogen isotopes is very 
small. The helium case is typified by many large contributions (in kHz) but
the scalar proton charge -- neutron magnetic moment term completely
dominates. The scalar neutron charge -- proton magnetic moment term is about
4\% of the dominant interaction, the tensor terms are 1-2\%, while the
convection current of the two protons is about 5\% of the dominant interaction,
making it the largest of the corrections.

Table~3 adds the results of Table~2 to the QED-only calculation for 1s states,
and expresses the differences with experiment as fractions of the Fermi energy.
Results must be considered quite good, given the size of our hadronic expansion
parameter. The deuterium case is particularly close to experiment, and this is 
likely due to the small binding energy, which tends to minimize relativistic 
corrections. The trinucleon cases range from very good in the $^3$H case ($\sim 
3\%$ residue) to adequate in the $^3$He case ($\sim 20\%$ residue). The large 
disparity in the two cases is undoubtedly due to missing MEC, particularly the 
isovector ones. Even this amount of missing strength is only slightly larger 
than our expansion parameter.

\begin{table}[htb]
\centering

\caption{Difference between hyperfine experiments and hyperfine calculations for
the 1s state of light hydrogenic atoms, expressed as parts per million of the
Fermi energy. The first line is the difference with respect to the QED
calculations only, while the second line incorporates the hadronic corrections
calculated above (the Zemach moment for hydrogen and nuclear corrections for 
the three nuclei).}
\vspace*{0.1in}
\begin{tabular}{|l || c c c c|}
\multicolumn{5}{c}{$(E_{\rm hfs}^{\rm exp} - E_{\rm hfs}^{\rm Th})/E_{\rm F}
\, {\rm (ppm)}$} \\ \noalign{\smallskip} \hline
Theory        & H & $^2$H & $^3$H & $^3$He$^+$ \rule{0in}{2.5ex}\\ \hline \hline
QED only      & $-$33     & 138   & $-$38 & $-$212       \\ \hline
QED + hadronic & 3.2(5) & $-$3.1   & 1.2   & $-$46        \\ \hline 
\end{tabular}
\end{table}

Previous work on this topic is quite
old\cite{d-th-1,d-th-0,t-th-1,t-th-2,he-th,d-t-he}, except for the
deuterium\cite{yulik} case. The older work relied on the Breit 
approximation for the electron physics, which is sufficient only for the
leading-order corrections. It used an adiabatic treatment of the nuclear
physics based on the Bohr picture of the nuclear hyperfine anomaly, which is far
more complex than the treatment that we have presented. Uncalculated QED
corrections and poorly known fundamental constants (such as $\alpha$) led to
estimates of nuclear effects that were many tens of ppm in error. Although the
nuclear physics at that time was not adequate to perform more than qualitative
treatments of the trinucleons, the SU(4) mechanism was known and this allowed a
qualitative understanding. The only previous attempt to treat the three nuclei
simultaneously was in Ref.~\cite{d-t-he}. They found nuclear corrections of
about 200 ppm for deuterium, 20 ppm for $^3$H, and $-$175 ppm for $^3$He$^+$.
Except for the deuterium case (which involves significant cancellations) this
has to regarded as quite successful, given the knowledge available at that time.

\section{Conclusions}

We have performed a calculation of the nuclear part of the hfs for $^2$H,
$^3$H, and $^3$He$^+$, based on an expansion parameter adopted from $\chi$PT,
a unified nuclear model, and modern second-generation nuclear forces. This 
is the first such calculation, and the results are quite good. Details of
the results can be understood in terms of the approximate SU(4) symmetry
that dominates the structure of light nuclei.

\section*{Acknowledgments}

The work of JLF was performed under the auspices of the U.\ S.\ Dept.\ of 
Energy, while the work of GLP was supported in part by the DOE.

\end{document}